\documentclass{piparticle-final}
\usepackage[T1]{fontenc}
\usepackage[latin9]{inputenc}
\usepackage{textcomp}
\usepackage{graphicx}

\usepackage{cite} % To compress citation ranges
\usepackage{epstopdf}

\begin{document}

\volume{5}               % To be inserted by Editor
\articlenumber{050005}   % To be inserted by Editor
\journalyear{2013}       % To be inserted by Editor
\editor{J. J. Niemela}   % To be inserted by Editor
\reviewers{V. Lakshminarayanan, Waterloo University, Canada}  % To be inserted by Editor
\received{7 December 2012}     % To be inserted by Editor
\accepted{19 June 2013}   % To be inserted by Editor
\runningauthor{P. Grondona \itshape{et al.}}  % To be inserted by Editor
\doi{050005}         % To be inserted by Editor

\title{Enhancement of photoacoustic detection of inhomogeneities in polymers}

\author{P. Grondona,\cite{inst1} H. O. Di Rocco,\cite{inst2}
D. I. Iriarte,\cite{inst2} J. A. Pomarico,\cite{inst2}\\
H. F. Ranea-Sandoval,\cite{inst2}\thanks{E-mail: hranea@exa.unicen.edu.ar} \hspace{0.5em}
G. M. Bilmes\cite{inst3}}

\pipabstract{We report a series of experiments on laser pulsed photoacoustic excitation
in turbid polymer samples addressed to evaluate the sound speed in
the samples and the presence of inhomogeneities in the bulk. We describe
a system which allows the direct measurement of the speed of the detected
waves by engraving the surface of the piece under study with a fiduciary
pattern of black lines. We also describe how this pattern helps to
enhance the sensitivity for the detection of an inhomogeneity in the
bulk. These two facts are useful for studies in soft matter systems
including, perhaps, biological samples. We have performed an experimental
analysis on Grilon\textsuperscript{\textregistered{}}samples in different
situations and we show the limitations of the method.
}

\maketitle

\blfootnote{
\begin{theaffiliation}{99}
   \institution{inst1} Universidad Nacional de Rosario. Facultad de Ciencias Bioquímicas
y Farmacéuticas. Rosario (Santa Fe) Argentina.
   \institution{inst2} Instituto de Física \textquotedblleft{}Arroyo Seco\textquotedblright{}, Universidad Nacional del Centro de la Provincia de Buenos Aires. Calle
Pinto 399, B7000GHG, Tandil (Buenos Aires) Argentina.
   \institution{inst3} Centro de Investigaciones Ópticas (CONICET-CIC) and Facultad de
Ingeniería Universidad Nacional de La Plata, La Plata. Argentina.
\end{theaffiliation}
}

\section{Introduction }

In highly light-scattering materials, such as certain types of polymers, turbid liquids, glassy structures, and body organs, inspection and
monitoring of internal features  were made possible by means of X-Ray
irradiation until the development of ultrasound imaging. The former
has the well-known disadvantage that in biological tissues it may
trigger degenerative processes in the cells, and in non-biological
samples, X-Ray inspection is not always simple to perform directly
in the production line. Ultrasound imaging is very helpful in these
situations. 

On the other hand, visible light optical tomography and optical topography
is nowadays reaching the status of clinical resource in the detection
and monitoring of several types of tumors and for non-invasive evaluation
of oxygenation of tissues in biological samples. In  non-clinical applications
it can be used for the detection of abnormal bodies within materials, which is of great importance in quality control in several areas of
technology. These techniques were derived from the study of light
propagating in turbid media, and applied afterward to biological samples
and medical imaging of different parameters often using polymers as
phantoms of biological tissues \cite{key-1,key-2,key-3,key-4,key-5,key-6,key-7}.

The photoacustic effect (PA) provides a method of analysis that has
been used in clear fluids and has sufficiently proven its capability
for detecting very low concentrations of absorbing species in a mixture
or solution; it has also been used for the monitoring of molecular
processes in different environments as shown in references \cite{key-7,key-8,key-9,key-10,key-11,key-12,key-13,key-14,key-15,key-16,key-17}. This paper intends to make a contribution on the
application of the PA in soft matter, namely the detection of inclusions
in polymer samples and the direct determination of the speed of sound
in the material used for the samples.

The PA technique has the advantage that acoustic waves do not scatter
as light does in the characteristic lengths of many experimental situations.
Even if the excitation light undergoes scattering, the location of
an inhomogeneity within the bulk of the sample can be achieved by
detecting the remnant of the shock wave generated at an absorbing
region or at an interface at which the speed of the sound waves changes.
Repeating this inspection at other relative positions of the laser
and the acoustic detector and with the aid of a suitable algorithm,
a sufficiently precise location of a single inhomogeneity of simple
geometry can thus, in principle, be resolved, together with some information
about its composition (using at least two wavelength for the excitation),
provided the speed of sound is known (see, for example, Ref. \cite{key-18}).
An example of this is presented in Ref. \cite{key-19-1} in a rear-detection
scheme used for detecting inhomogeneities in subsurface inhomogeneities
in metals.

The PA detection of bodies included in a turbid medium may provide
complementary information to diffuse light propagation studies in
that medium. Namely, it could bring an independent value for the absorption
coefficient, and it thus may help in the solution of the inverse problem
in optical tomography of samples. 

The speed of sound determination relies on the fact that the acoustic
signal picked by the transducer arrives at times proportional to the
distance from the laser beam that generates the shock wave to that
transducer. A drawback with the photoacustic method applied to turbid
materials is that the light scattered by the bulk generates a pressure
pulse on the detector if it is in contact with a free surface of the
sample explored. Consequently, the time of arrival of the pressure
signal at the detector is insensitive to the relative position of
the laser and the sensor. Hence, the speed of the waves involved in
the PA signal is difficult to determine and requires an adequate  procedure
to evaluate it. This is one of the motivations of this contribution.

 For this paper, we used a laser pulsed photoacustic system equipped
with a PZT in contact with the sample made of polymer Grilon\textsuperscript{\textregistered{}}which
is representative of a turbid medium, to show how the presence of
controlled, fiduciary absorbing regions in the surface of a sample
are used as local wave generators that allow the determination of
the speed of waves in materials despite of the light scattering described.
We have engraved in the surface of the samples a pattern of stripes
of absorbing material. In this way, we have a greater signal whose
contribution may be discriminated from the signal generated by the
light scattered by the sample material. We demonstrate that this fiduciary
pattern is useful also to enhance the photoacustic signal, and that
from that signal the presence of inhomogeneities in a medium may be
inferred.

Other successful recent approaches to the problem of detection of
tumoral tissues in biological samples can be found in  Refs.
\cite{key-17.1,key-17.2}.

\section{Experimental}

\begin{figure}[h]
%[bb=4bp -1bp 842bp 592bp,scale=0.25]
\centering{}\includegraphics[width=\columnwidth]{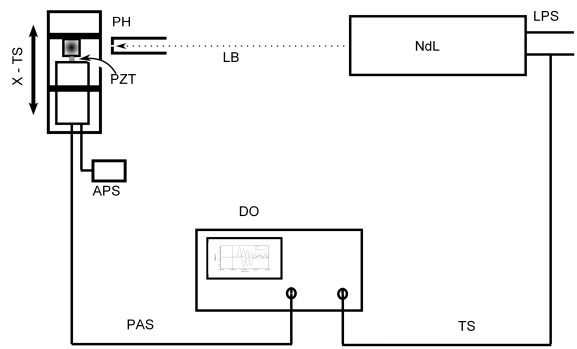}\caption{Experimental setup of the experiment.  The parts are: the Nd$^{+3}$:YAG
laser (NdL), the positioning device (X-TS), the Pinhole (PH) to clip
the laser beam (LB), and the oscilloscope (DO), the amplifier with
a DC power supply (APS). The DO is synchronized with the laser via
the laser pulse synchronization (LPS) cable.\label{fig: Figure 1}}
\end{figure}

A scheme of the pulsed photoacustic system used in all the experiments
is shown in Fig. \ref{fig: Figure 1}, which is essentially the
same that can be used to determine the speed of sound in liquids and
in clear samples. 

We used a pulsed Nd\textsuperscript{+3}:YAG laser emitting at $1.06\,\mu$m
with a pulse duration of approximately $10$ ns, at energies between
$0.5$ mJ to 50 mJ. The laser beam was clipped by means of a
pinhole in order to reduce the original laser beam size and to use
a uniform spot thus reducing the power impinging on the samples. This
pinhole was held at the far end of a beam dump for security reasons.
In the results we present here, we have used two pinholes of $1$ mm
and $1.5$ mm in diameter which shall be specified in each experiment.
This is the diameter of the laser impinging on the sample, as inferred
from sensitive photographic paper. For acoustic detection, a ceramic
$4\times4$ mm$^{2}$ PZT transducer was strongly pressed against one
of the free surfaces of the sample, namely the one normal to that
facing the laser. The photoacustic signals were amplified and processed
by means of a Tektronix TDS 3032B, $300$ MHz digital oscilloscope,
averaging at least 64 signals before displaying the photoacustic signal. 

Samples used were square-section parallelepipeds, $10$ mm width,
and $39$ mm high, all made from the same polymer Grilon\textsuperscript{\textregistered{}}piece.
The samples were placed in a C-clamp, with the PZT cage in one of
its arms. The fiduciary pattern engraved on one of the faces of some
samples consists of five grooves of approximately $1$ mm width
and $0.2$ mm deep, filled with thick black paint, separated by
stripes of material which  retain the natural turbid white color of
the polymer (which we call \textquotedblleft{}clear\textquotedblright{}
for short) of $1$ mm, whose lengths are approximately 70\% the
length of the face. A second type of sample prepared in a similar
fashion, but with a centered cylindrical hole of 3 mm diameter drilled
in it parallel to the surfaces of the sample in all cases mentioned,
was also used in the experiments in order to compare the signals with
the former. This cavity was alternatively emptied or filled with deionized
water. We call ``sample 1'' the one drilled with the cylindrical cavity,
and ``sample 2'' the one without the hole. Figure \ref{fig: Figure 2}
is a sketch of sample 1 with a schematic representation of the fiduciary
pattern used. The PZT and the laser beam relative positions are displayed,
together with the approximate position of the cavity. 

\begin{center}
\begin{figure}[h]
\begin{centering}
\includegraphics[scale=0.25]{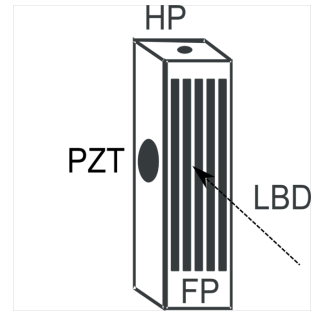}
\par\end{centering}

\caption{The Grilon\textsuperscript{\textregistered{}}sample prepared for
the surface absorption experiments. The shadowed region (PZT) is the
location of the transducer with respect to the impinging laser beam
direction (LBD). The cavity is a $3$ mm diameter hole, whose position
(HP) is shown for the samples that have drilled cavities. The cavity
may be empty or filled with water. The height of the samples is $39$ mm
and has a $10$ mm square base. It has five grooves (FP) in its
front face, painted in black to enhance absorption. \label{fig: Figure 2}}
\end{figure}

\par\end{center}

We obtained two types of signals, those from samples without holes
and those from samples with centered holes. Each type was subdivided
into signals taken with the laser impinging on the blank surface,
and those taken with the laser impinging on the patterned surface.
Besides, there are signals obtained from the samples with cavities,
either empty or filled with water. 

In each sample, the laser point of impact was moved from the farthest
possible position to the nearest with respect to the PZT. This was
accomplished by means of a $1\,\mu$m precision, step motor movable
stage, Zaber Model T-LA60A, controlled by a PC interface.

\section{Results }

\begin{figure}[h]
\begin{centering}
\includegraphics[width=\columnwidth]{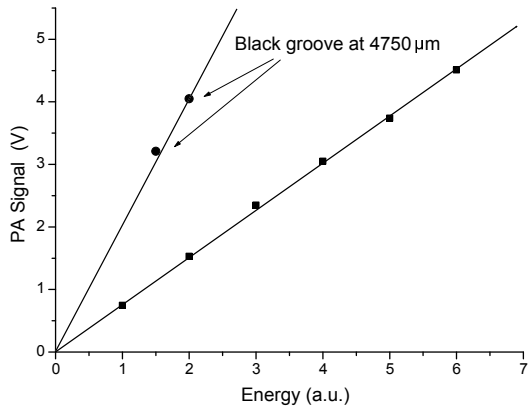}
\par\end{centering}

\caption{PAS vs. laser pulse energy. The linearity of the PA response (squares)
to laser excitation is evident in the plots. The black dots represent
the PA signal when the laser impinges on a black stripe nearly at
the center of the front face of a striped sample. \label{fig: Figure 3}}
\end{figure}

In order to properly analyze the results, we calibrate the response
of the system to increasing laser pulse energy. To this end we irradiate
a blank surface of sample 2 at a point near the center of the face,
and we plotted the amplitude of the first maximum of the acoustic
signal as a function of the laser pulse energy. The result is displayed
in Fig. \ref{fig: Figure 3} and shows linearity in the energy range
used.

In the same plot, we display three points (including the origin) which
are the maxima of the signal at the same location of a striped sample
face, but impinging on a black groove. As it can be seen, the signal
nearly trebles its maximum peak for the same excitation energy. In
both experiments, the pinhole used was $1.5$ mm in diameter. 

\begin{figure}[h]
\begin{centering}
\includegraphics[width=\columnwidth]{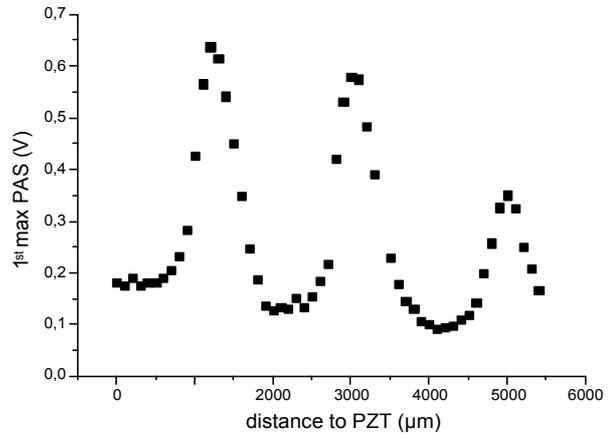}
\par\end{centering}

\caption{The centers of clear bands and black grooves are clearly resolved
by scanning the surface with the laser. The signals were taken at
$100\,\mu$m displacement from each other. The energy of PAS decreases
with distance of the laser beam to PZT. The increment in signal due
to the grooves is more than 6-fold with respect to the signal due
to the bulk polymer.\label{fig: Figure 4}}
\end{figure}

After ascertaining the linearity of the response, and the fact that
there is an evident dependence of the PAS on the absorbance of the
surface, we obtain a profile of three of the grooves of sample 2 by
plotting the value of the amplitude of the first peak of the PA signal
versus the relative distance between the PZT and the excited region
using the same pinhole as before. The result of this is shown in Fig.
\ref{fig: Figure 4}. It can be seen that 1) the groove profile is
neatly resolved, and 2) there is an improvement of the signal generated
in the black stripes which decreases as the distance increases. Since
the stripes and the laser beam have approximately the same transverse
size of the grooves, the resulting profile is somehow rounded off,
but this is not important in what we  aim to prove here.

We could determine the speed of sound in the sample from a plot of
the time position of the beginning of the first peak of the acoustic
signal (arrival time), as a function of the distance between the impinging
point on the sample and the PZT detector. But when we try to do that,
experiments  demonstrate that the time elapsed since a laser pulse
triggers a digital oscilloscope and the appearance of the PA signal
is the same regardless of the distance between the impinging laser
beam and the detector, due mainly to the light scattered by the bulk
of the polymer that hits the PZT. This poses a problem in the evaluation
of the speed of the waves. 

To avoid that difficulty, we use the signal produced if the laser hits
in the black  grooves,  generated only by the absorption at the  grooves.
We obtain this by subtracting from the PAS signal measured when the
laser impinges in a black  groove, the PAS signal obtained in a \textquotedblleft{}clear\textquotedblright{}
region nearby. To this end, we moved the sample slightly away from
the previous black stripe, so the first PAS signal was obtained with
the beam impinging in a black  groove and the second PAS signal was
obtained with the beam impinging in a white stripe. 

\begin{figure}[h]
\begin{centering}
\includegraphics[width=\columnwidth]{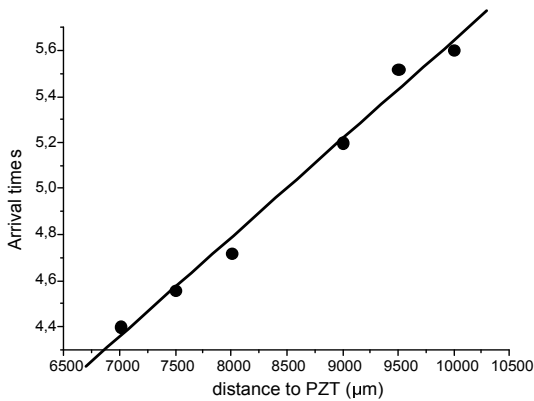}
\par\end{centering}

\caption{Time of the first maximum of the processed signals vs. the position
of the impinging point of the laser presents a linear correlation
($R\,\approx\,0.994$) that yields a value of $v=(2333\,\pm\,133)$ ms$^{-1}$
for the speed of sound. All these signals were taken with a pinhole
of $1$ mm diameter. \label{fig: Figure 5}}
\end{figure}

Figure \ref{fig: Figure 5} shows the determination of the speed of
sound in sample 2 by this method. Since the plot uses as input the
maxima of the amplitude of the signals, the straight line would not
cross the origin. The extrapolated value for zero-crossing corresponds
approximately to the amplitude of the first maximum of the signal
in clear samples. 

Evaluation of the slope of the resulting line allows the calculation
of the value of speed, as $v=(2333\pm133)$ ms$^{-1}$, which compares
well with calculated data determined by using the properties of the
polymer \cite{key-19}. Since the vibration of the whole sample
has distinctive frequencies, the FFT of the PA signal provides another
estimation of the speed of waves, once the frequency sequence is properly
found. The Fast Fourier Transform (FFT) analysis was used both, to
estimate the sound speed in the Grilon\textsuperscript{\textregistered{}}sample
via the frequencies identified in the spectrum and their spacing,
knowing that the piece is a parallelepiped of known dimensions, and
to define a scale for the energy of the pulse. The details of this
procedure are straightforward  calculations \cite{key-20}. We found
it useful to use the power spectrum for defining the energy instead
of the integral of the temporal pulse and that is the parameter we
use in the presentation of the results. 

\begin{figure}[h]
\begin{centering}
\includegraphics[width=\columnwidth]{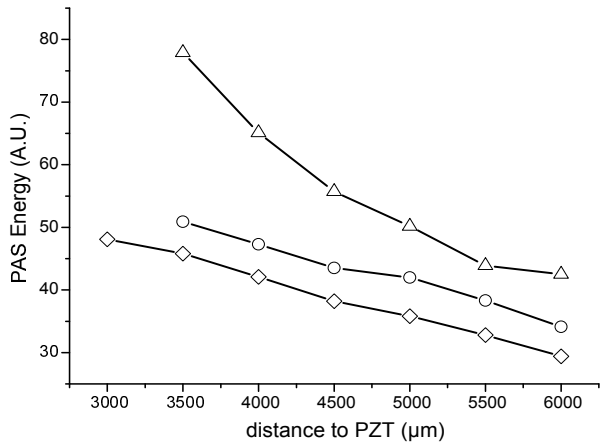}
\par\end{centering}

\caption{The acoustic energy (FFT power integral of the PAS) vs. the relative
distance between the laser beam and the PZT in Grilon\textsuperscript{\textregistered{}}samples
in a face with no fiduciary pattern (clear sample). The energy diminishes
as the distance to the detector increases. References in the insert:
Triangles represent the cavity empty in a clear sample. Circles are
for cavity filled with water in clear samples. Rhombi are for clear
samples without the cavity.\label{fig: Figure 6}}
\end{figure}

\begin{figure}[h]
\includegraphics[width=1.2\columnwidth]{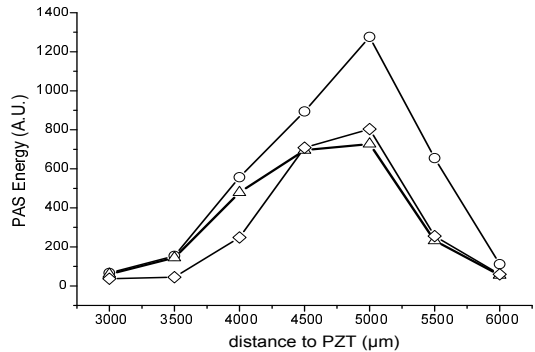}
\caption{The energy deposited by the laser on a white stripe and on the black
grooves for the three cases analyzed as seen in the insert. The circles
represent a grooved sample with the cavity filled with water. Triangles
and rhombi are for grooved samples and the empty cavity, and no cavity,
respectively.\label{fig: Figure 7}}
\end{figure}

Figures \ref{fig: Figure 6} and \ref{fig: Figure 7} show an FFT treatment
of the signals obtained from the following three cases: solid Grilon\textsuperscript{\textregistered{}}sample,
sample with empty cavity and sample with the hole filled with deionized
water. The PAS energy used in this figure is a measure of the energy
content of the acoustic pulse, as evaluated from the power spectrum
of the signal. 

In Fig. \ref{fig: Figure 6}, we display the results obtained impinging
with the laser on clear faces, and in Fig. \ref{fig: Figure 7}
we show the results impinging with the laser on the patterned faces.
It is clear that a distinctive feature arises near the center of the
sample in the patterned faces where a black stripe is located, which
is not visible in the clear-face analysis.

The acoustic energy deposited in the patterned faces is more than
one order of magnitude higher than that obtained in the clear faces
when the inclusion is present (compare Fig. 6 with Fig. 7). The
water-filled hole and the empty hole are also clearly distinguishable
from the solid Grilon\textsuperscript{\textregistered{}}response.

\section{Analysis and Conclusions }

We have performed an experimental analysis of the photoacustic signal
in Grilon\textsuperscript{\textregistered{}}polymer but it can be
extended to other materials as  epoxy resins used also as biological
phantoms. We have shown that such applications are viable
for quantitative determinations. The above results can be used to
determine the presence and some optical characteristics of an inhomogeneity
embedded in this type of materials. 

All the acoustic signals detected by the PZT in this soft turbid solid
material begin at approximately the same time after the laser trigger
fires, regardless of the relative distance from the impact point of
the excitation to the PZT acoustic detector, due mainly to scattering,
making this method useless to evaluate the speed of sound by the scanning
standard procedure. 

The differences in those PA signals are difficult to analyze. Therefore, to evaluate the speed of the acoustic waves and to gather information
about the presence of a cavity or inhomogeneity in the polymer, we
have developed a method that used a regular pattern on the surface
of the sample, consisting on parallel clear stripes of the base material
and grooves filled with highly absorbent black paint. In the black
grooves, the localized absorption provides a strong shock wave at
the surface. By comparing the time of appearance of the signal
in different positions of the surface, it is possible to estimate
the speed of those waves in the polymer. 

For each of the zones in the pattern, the PA curves undergo a change
of shape and amplitude from signals in the white zones to the signals
obtained in the black stripes, being this strong evidence of the effect
of the inhomogeneity in the signal. A Power FFT was performed on each
signal in order to provide another means to determine whether a black
or a white stripe is excited, and the integral of the FFT provides
a measure of the energy absorbed by the sample in each case. Please
note that the energy of the laser was fixed to a value that avoids
bleaching of the paint, being in all cases below $500\,\mu$J per pulse.

The signals generated at these inhomogeneities provide a well defined
point of absorption and thus a definite path for the sound generated
by the light-absorption mechanism which is very distinctive from other
mechanisms of excitation of the PZT. It also reveals the presence
of a surface inhomogeneity once the contributions of other sources
of acoustic waves are identified making suitable use of reference
signals. All the conclusions of this work must be under the proviso
that the PZT has a limited frequency band. 

Although the above results were obtained for a soft polymer with a
fiduciary painted pattern, they can be extended to other type of resins
with charge of dyes or other absorbent particles. We are confident
that with minor modifications it can be used for the determination
of properties of materials of biological interest as well. 

The results shown in Figs. \ref{fig: Figure 6} and \ref{fig: Figure 7}
confirm that employing one of the surfaces of the sample conveniently
patterned, and scanning it for detection of ultrasound signals, can
be used to determine the presence of an inhomogeneity, albeit its
precise location and size is not well defined by this procedure and
it should be complemented by similar determinations at other relative
positions of the laser and the PZT. The increase in the signal with
respect to the background material is at least one order of magnitude
or better.

 When using this technique in phantoms used in medical applications,
one should take care of the fact that there are limitations in several
aspects, such as the power involved in each pulse avoiding any kind
of damage, and that using other wavelengths would be better suited
for biological tissues which involve blood. Other type of samples
are being currently inspected by modifications of the procedure reported
here so to adapt it to gelled phantoms.

The conclusions are, in short, that the system is sensitive to the
presence of the inhomogeneity, and that the higher absorbance of the
painted stripes in the surface allows not only to evaluate the speed
of sound (which is essential to any tomographic technique) but also improves
 the detectivity by enhancing the energy released as mechanical
waves. This is a non-trivial result since in the modeling of the propagation
of the laser light in the turbid substance, scattering is predominant,
but still is sufficient for the detection of inhomogeneities through
changes in the absorption. The technique based on the PA is simple
and has the advantage that it can be adapted to be used in larger
samples or in samples of biological interest. The procedure of using
a single acoustic detector for the signals produced by the laser scanning
of the surface under study, has an advantage over multiple detector
arrangements in the sense that with a suitable fiduciary pattern the
method can provide information about the speed of the waves involved
in the signal. This is interesting because the data processing would
not depend on generic information about its value. 

\begin{acknowledgements}
PG wants to thank the Red Nacional de Laboratorios de Óptica for financial
help and partial funding during the experiments and to InterU System
for providing a grant for the completion of the experiments. 
This work partially funded by Universidad Nacional del Centro de la
Provincia de Buenos Aires, 
Agencia Nacional de Promoción Científica y Tecnológica (PICT 0570) and
CONICET (PIP 384). 
HODR, DII, JAP and HFRS are members of  Carrera
del Investigador Científico, Consejo Nacional de Investigaciones Científicas
y Técnicas (Argentina). 
GMB is member of  Carrera del Investigador Científico, Comisión
de Investigaciones Científicas de la Provincia de Buenos Aires (Argentina). 
Authors wish to thank Nicolás A. Carbone for help in the final
preparation of the  manuscript.
\end{acknowledgements}

\end{document}